\documentclass[12pt]{article}
\usepackage{graphicx}
\usepackage[left=2cm,right=2cm,top=2cm,bottom=2cm]{geometry}
\usepackage{xcolor}
\usepackage{caption}
\usepackage{subcaption}
\usepackage{lineno}
\usepackage[
    colorlinks=true,
    linkcolor=blue,
    filecolor=magenta,      
    urlcolor=blue,
    pdftitle={Overleaf Example},
    pdfpagemode=UseNone,
    ]{hyperref}

\newcommand\pubdate{\today}

\def\Title#1{\begin{center} {\Large #1 } \end{center}}
\def\Author#1#2{\begin{center}{ {\sc #1}\\{#2}} \end{center}}
\def\Address#1{\begin{center}{ \it #1} \end{center}}

\newcommand\pubblock{\rightline{\begin{tabular}{l}  \\ 
         \pubdate  \end{tabular}}}
\newenvironment{Abstract}{\begin{quotation}  }{\end{quotation}}
\newenvironment{Presented}{\begin{quotation} \begin{center} 
             PRESENTED AT\end{center}\bigskip 
      \begin{center}\begin{large}}{\end{large}\end{center} \end{quotation}}

\begin{document}
\begin{titlepage}
 \pubblock
\vfill
\Title{Measurements of top-quark production cross sections with the ATLAS detector}
\vfill
\Author{M. A. Principe Martin,}{On behalf of the ATLAS Collaboration}
\Address{Universidad Autónoma de Madrid (Spain)}
\vfill
\begin{Abstract}
The Large Hadron Collider (LHC) produces a vast sample of top-quark pairs and single-top quarks. Measurements of the inclusive top-quark production rates at the LHC have reached a precision of several percent and test advanced next-to-next-to-leading-order predictions in QCD. Measurements of production cross sections test the Standard Model predictions and help to improve the Monte Carlo models. In this contribution, comprehensive measurements of top-quark-antiquark pair and single-top quark production are presented; the measurements use the data recorded by the ATLAS experiment in the years 2015-2018 during Run 2 of the LHC. A recent result from the 5 TeV operation of the LHC is also included. In addition, a first look into top-quark pair production in Run 3 data at 13.6 TeV is also presented.
\end{Abstract}
\vfill
\begin{Presented}
DIS2023: XXX International Workshop on Deep-Inelastic Scattering and
Related Subjects, \\
Michigan State University, USA, 27-31 March 2023 \\
     \includegraphics[width=9cm]{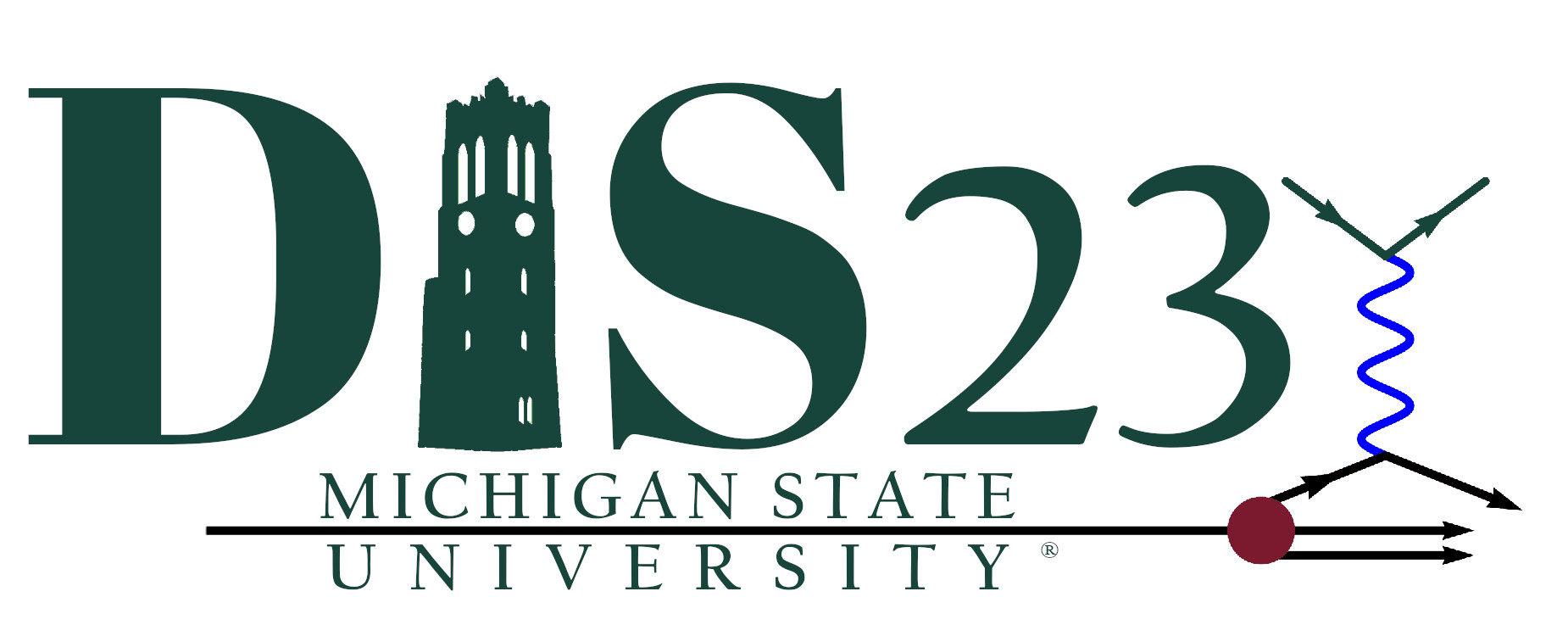}
\end{Presented}
\vfill
\hrule
\vspace{5pt}{\footnotesize\raisebox{0.5ex}{\scriptsize \textcopyright} 2023 CERN for the benefit of the ATLAS Collaboration.\\
Reproduction of this article or parts of it is allowed as specified in the CC-BY-4.0 license.}
\end{titlepage}

\textbf{1. Introduction}\medskip

Top quarks are predominantly produced in $pp$ collisions in pairs via QCD and singly via electroweak (EW) interactions. The Large Hadron Collider (LHC) can be considered as a top factory, which provides high-statistics data samples used to test the Standard Model (SM) and search for new phenomena. Electroweak tests are done using single-top quark production while measurements of $t\bar{t}$ production allow tests of QCD at the highest accessible energy scales. For the $t\bar{t}$ production, next-to-next-to-leading-order plus next-to-next-to-leading-logarithm (NNLO+NNLL) predictions are available and the measurements have been used to constrain the Parton Distribution Functions (PDF) in global fits. Measuring with precision top-quark production is a key factor for Beyond Standard Model (BSM) searches as it is the main background.\medskip

The production of single-top quarks proceeds via three channels: $t$- and $s$-channels, when the top quark is produced by a $W$-boson in these channels, and the $Wt$ channel, when the top quark is produced in association with a $W$-boson. The main backgrounds for single-top quark production are $t\bar{t}$ and $W$+jets processes. Other background processes are $Z$+jets, diboson and multijet production.\medskip

Top-pair production is usually classified according to the products of the $W$-decays in all-hadronic, semileptonic and dileptonic channels. Single-top and $W$+jets are the main background processes; for the all-hadronic channel, multijet background is also important.\medskip

Recently, the ATLAS~[\hyperlink{bib:ATLAS}{1}] Collaboration published several results measuring the top-quark production cross section both in pairs and singly.\medskip

\textbf{2. Single-top cross section}\medskip

A measurement of the single-top quark production cross section in the $s$-channel in $pp$ collisions at $\sqrt{s} = 13$~TeV with the ATLAS detector~[\hyperlink{bib:2209.08990}{2}] was performed. Previous measurements achieved significances of $2.5\sigma$ at $\sqrt{s}=7$ and $8$~TeV by CMS~[\hyperlink{bib:1603.02555}{3}], and $3.2\sigma$ at $\sqrt{s}=8$~TeV by ATLAS~[\hyperlink{bib:1511.05980}{4}]. The new measurement at $\sqrt{s} = 13$~TeV using $139$~fb$^{-1}$ was performed by selecting a charged isolated lepton (electron or muon), large $E_T^{miss}$ and two $b$-tagged jets. The method used to extract the signal is based on matrix element calculations to compute a discriminant that assigns to each event the probability of being a signal or background process. The production cross section was measured using a binned profile maximum-likelihood fit of the discriminant distribution. The measured single-top cross section is $\sigma=8.2^{+3.5}_{-2.9}$~pb with a significance of the signal over the background-only hypothesis of $3.3\sigma$. This result is in agreement with the theoretically predicted cross section computed at next-to-leading-order (NLO) with a value of $\sigma^{\rm SM}=10.3\pm0.4$~pb and an expected significance of $3.9\sigma$.\medskip

\textbf{3. Top-pair cross section}\medskip

Three new ATLAS measurements and one combination of ATLAS and CMS results were recently published.\medskip

A measurement at $\sqrt{s}=5.02$~TeV, using $\mathcal{L}=257$~pb$^{-1}$ of ATLAS data~[\hyperlink{bib:2207.01354}{5}], combines the semileptonic and dileptonic channels. The dileptonic channel uses the event counts while the semileptonic uses a boosted decision tree to increase the separation of signal and background processes. The combined measured cross section is $\sigma=67.5\pm2.7$~pb which agrees with the NNLO+NNLL QCD prediction of $\sigma^{\rm SM}=68.2^{+5.2}_{-5.3}$~pb.\medskip

For the measurement at $\sqrt{s}=7$~TeV, using $\mathcal{L}=4.6$~fb$^{-1}$ of ATLAS data~[\hyperlink{bib:2212.00571}{6}], a support vector machine method was applied in the semileptonic channel to separate signal and background processes. The measured cross section is $\sigma=168.5^{+7.1}_{-6.7}$~pb that is in agreement with the NNLO+NNLL QCD calculation of $\sigma^{\rm SM}=177^{+10}_{-11}$~pb.\medskip

The cross section measured by ATLAS using the data taken during 2022 at $\sqrt{s}=13.6$~TeV with a $\mathcal{L}=11.3$~fb$^{-1}$~[\hyperlink{bib:CONF}{7}] is $\sigma=859\pm29$~pb, in agreement with the NNLO+NNLL QCD prediction of $\sigma^{\rm SM}=924^{+32}_{-40}$~pb. The measurement was performed using the event-count method in the dileptonic channel and selecting opposite sign and different flavour leptons in the final state.\medskip

A combined measurement of the top-pair cross section was performed with ATLAS and CMS data using an integrated luminosity of $5$~fb$^{-1}$ at $\sqrt{s}=7$~TeV and $20$~fb$^{-1}$ at $\sqrt{s}=8$~TeV~[\hyperlink{bib:2205.13830}{8}]. The measurement was done using the decays with an opposite-charge $e\mu$ pair in the final state. The result of the measurement at $\sqrt{s}=7$~TeV is $\sigma\left(7\right.$~TeV$\left.\right)=178.5\pm4.7$~pb, in agreement with the NNLO+NNLL QCD prediction of $\sigma^{\rm SM}=177^{+10}_{-11}$~pb. In the case of the cross section at $\sqrt{s}=8$~TeV, the measurement is $\sigma\left(8\right.$~TeV$\left.\right)=243.3^{+6.0}_{-5.9}$~pb  and the NNLO+NNLL QCD calculation is $\sigma^{\rm SM}=255.3^{+10.6}_{-12.2}$~pb. A measurement of the ratio of the cross section at $\sqrt{s}=8$~TeV and the cross section at $\sqrt{s}=7$~TeV was also performed, which yields a result of $R_{8/7}=1.363\pm0.032$, which is in agreement with the prediction of $R_{\rm SM}=1.428^{+0.005}_{-0.004}$. In addition, fits to the combined measurement were performed to extract $m_t^{\rm pole}$ and $\alpha_{\rm s}\left(m_Z\right)$. The results are $m_t^{\rm pole}=173.4^{+1.8}_{-2.0}$~GeV (with $\alpha_{\rm s}\left(m_Z\right)$ fixed to $0.118\pm0.001$) and $\alpha_{\rm s}\left(m_Z\right)=0.1170^{+0.0021}_{-0.0018}$ (with $m_t^{\rm pole}$ fixed to $172.5\pm1.0$~GeV).\medskip

\textbf{4. Differential cross sections for $t\bar{t}$ production}\medskip

\begin{figure}[t]
     \centering
     \begin{subfigure}[b]{0.40\textwidth}
         \centering
         \includegraphics[width=\textwidth]{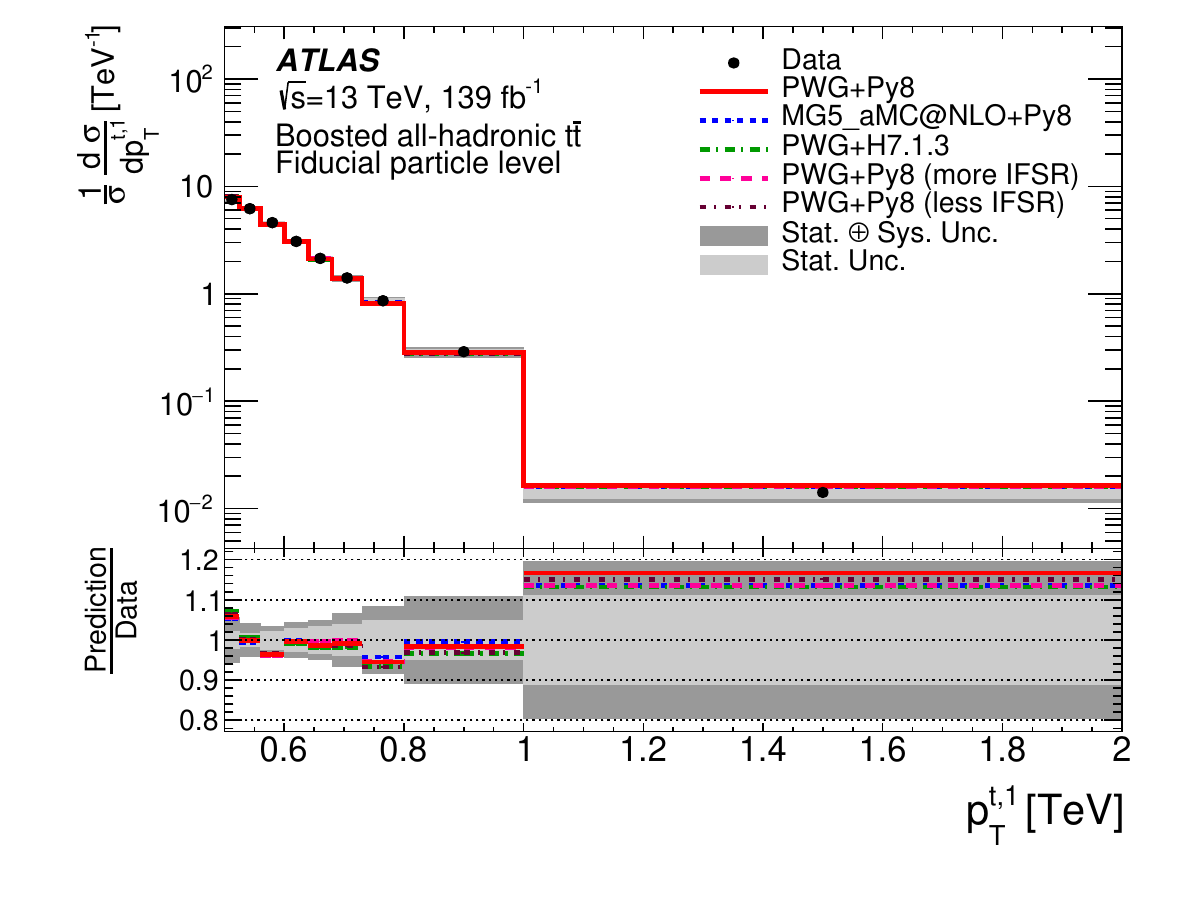} 
         \caption{}
     \end{subfigure}
     \begin{subfigure}[b]{0.50\textwidth}
         \centering
         \includegraphics[width=\textwidth]{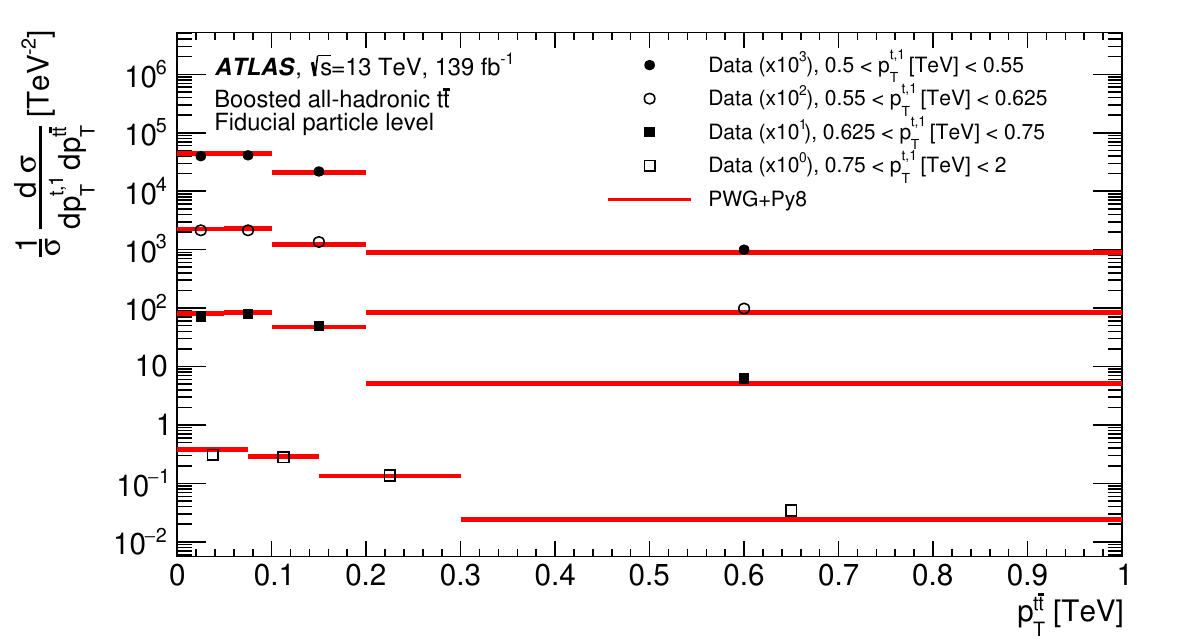}
         \vspace{0pt}
         \caption{}
     \end{subfigure}
        \caption{\small (a) Single normalised differential cross section at particle level as a function of the transverse momentum of the leading top quark. (b) Double normalised differential cross section at particle level as a function of the transverse momentum of the $t\bar{t}$ system in different regions of the transverse momentum of the leading top quark~[\protect\hyperlink{bib:JHEP04}{9}]. In these figures, the data are represented by markers and several NLO QCD predictions are presented as lines in different colours.}
        \label{fig:differential all-hadronic particle}
\end{figure}

\begin{figure}[t]
     \centering
     \begin{subfigure}[b]{0.49\textwidth}
         \centering
         \includegraphics[width=\textwidth]{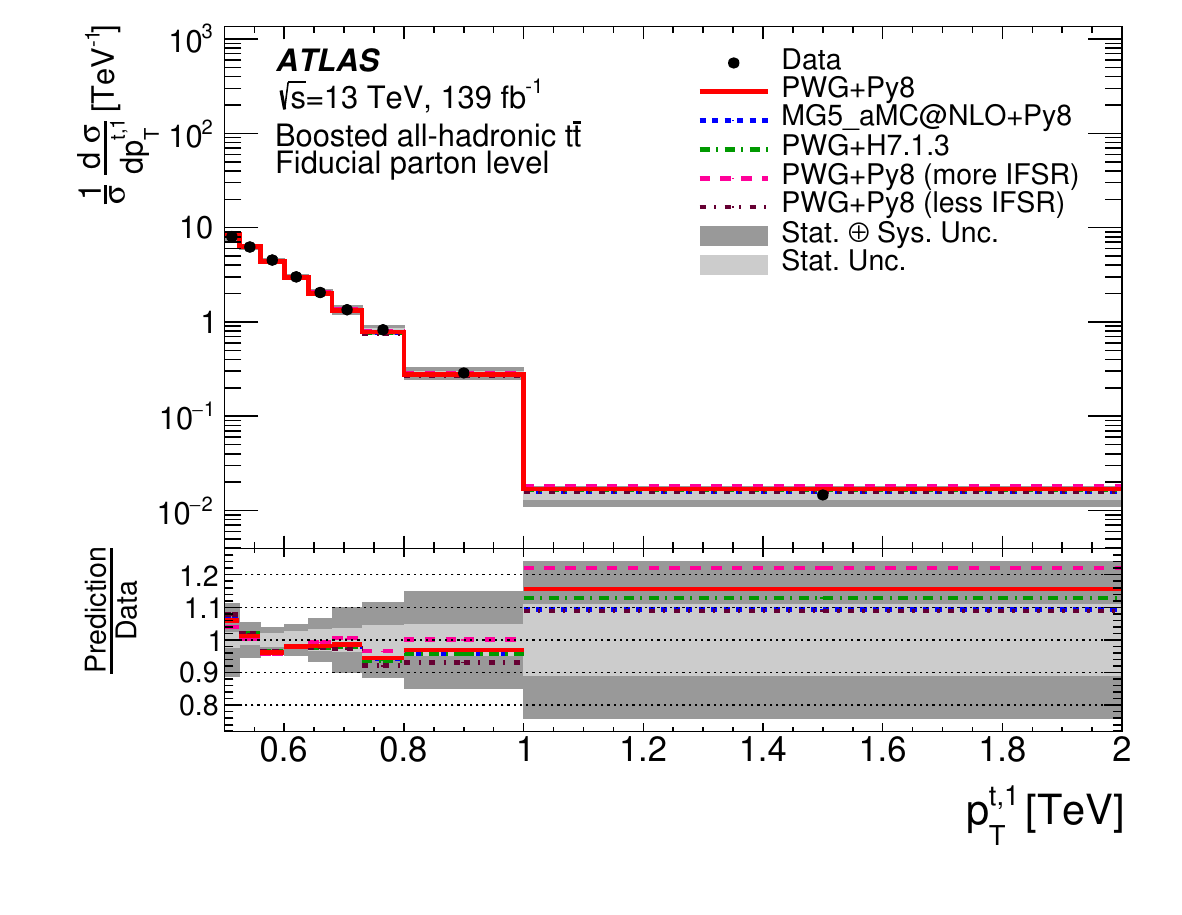} 
         \caption{}
     \end{subfigure}
     \begin{subfigure}[b]{0.49\textwidth}
         \centering
         \includegraphics[width=\textwidth]{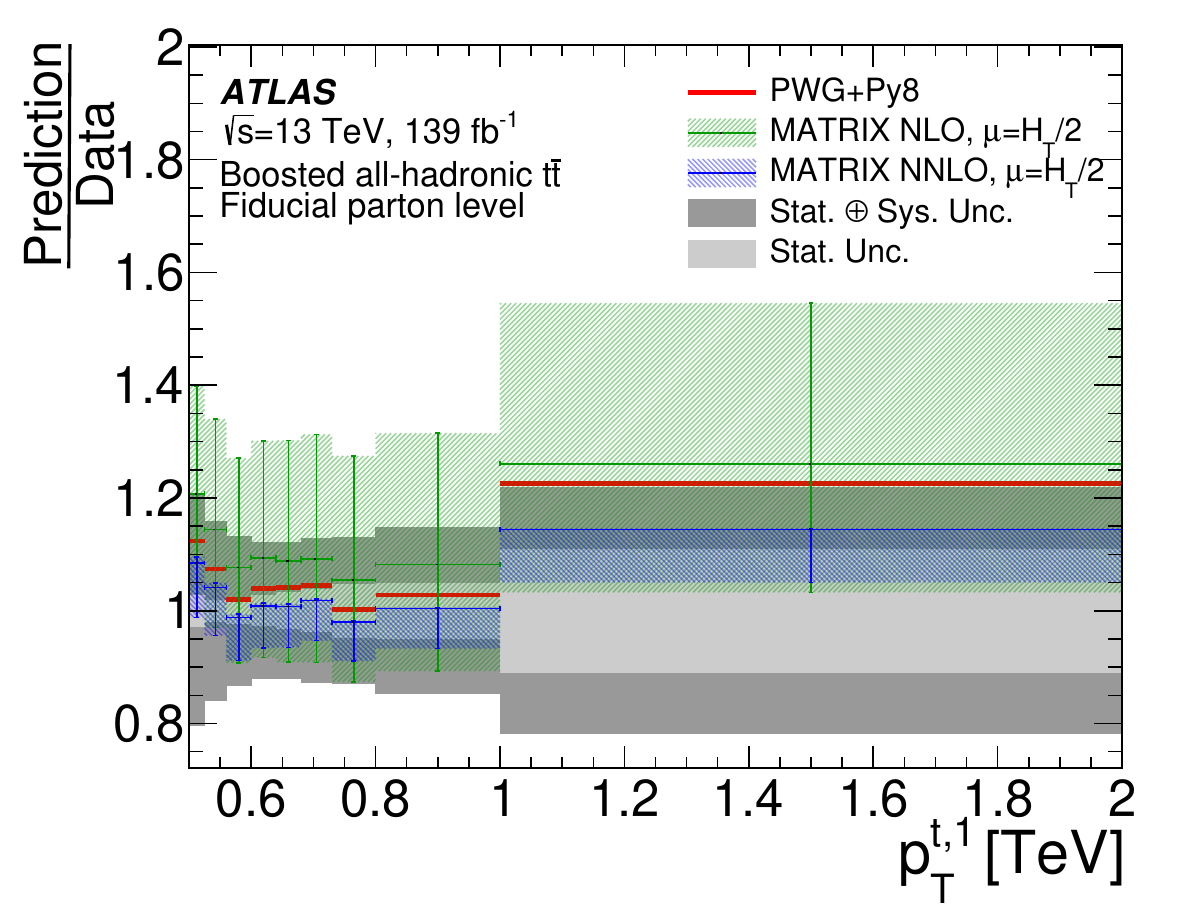} 
         \caption{}
     \end{subfigure}
        \caption{\small (a) Single normalised differential cross section at parton level as a function of the transverse momentum of the leading top quark; the data are represented as dots and several NLO QCD predictions are presented as lines in different colours. (b) Ratio of the predicted and the measured differential normalised cross section as a function of the transverse momentum of the leading top quark for the NLO \textsc{Powheg}+\textsc{Pythia}8 prediction as a red solid line and NLO (NNLO) \textsc{Matrix} prediction in green (blue)~[\protect\hyperlink{bib:JHEP04}{9}].}
        \label{fig:differential all-hadronic parton}
\end{figure}

Differential cross sections were measured in the all-hadronic final state using boosted top quarks with $\mathcal{L}=139$~fb$^{-1}$ at $\sqrt{s} = 13$~TeV~[\hyperlink{bib:JHEP04}{9}]. This analysis used Deep Neural Networks for the signal extraction. The selection requires the leading (subleading) top-quark transverse momentum to be greater than $500$~($300$)~GeV. At particle level, measurements of normalised single and double differential cross sections were performed and compared with NLO QCD predictions. Figure \ref{fig:differential all-hadronic particle} shows the normalised single differential cross section as a function of the transverse momentum of the leading top quark and the normalised double differential cross section as a function of the transverse momentum of the $t\bar{t}$ system in different regions of the transverse momentum of the leading top quark. Normalised differential cross sections were also measured at parton level. Figure \ref{fig:differential all-hadronic parton} shows the cross section as a function of the transverse momentum of the leading top quark. The NLO predictions show disagreements with the data, whereas the NNLO predictions give an improved description of the data. This analysis includes an Effective Field Theory (EFT) interpretation performed using dim6top~[\hyperlink{bib:dim6top}{10}] and EFT\textit{fitter}~[\hyperlink{bib:EFTfitter}{11}]. Seven Wilson coefficients in the Warsaw basis were individually fitted using the transverse momentum of the leading top-quark distribution. Quadratic terms were included in the fits, which lead to tighter bounds on the Wilson coefficients.\medskip

\begin{figure}[t]
     \centering
     \begin{subfigure}[b]{0.32\textwidth}
         \centering
         \includegraphics[width=\textwidth]{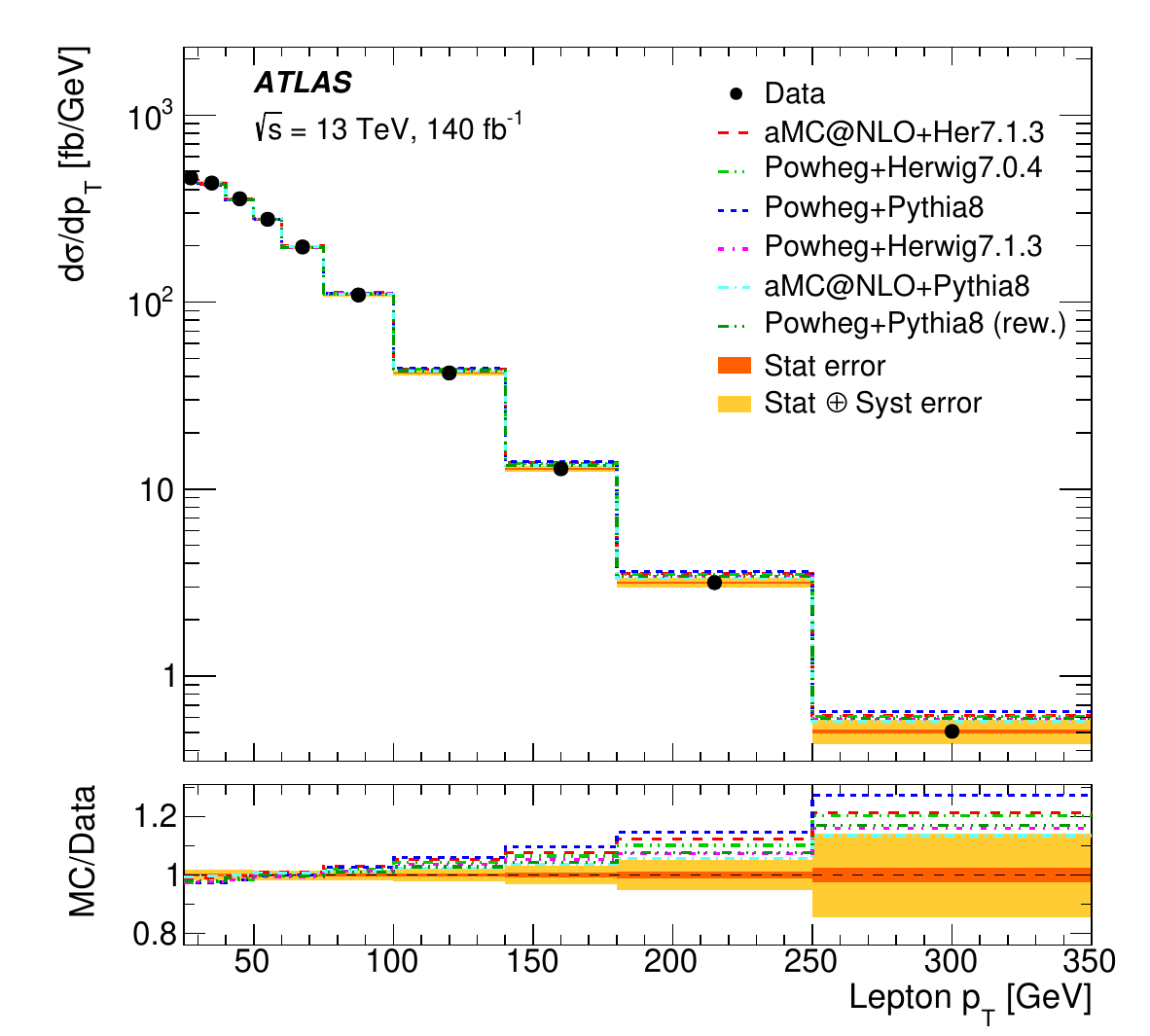} 
         \caption{}
     \end{subfigure}
     \begin{subfigure}[b]{0.32\textwidth}
         \centering
         \includegraphics[width=\textwidth]{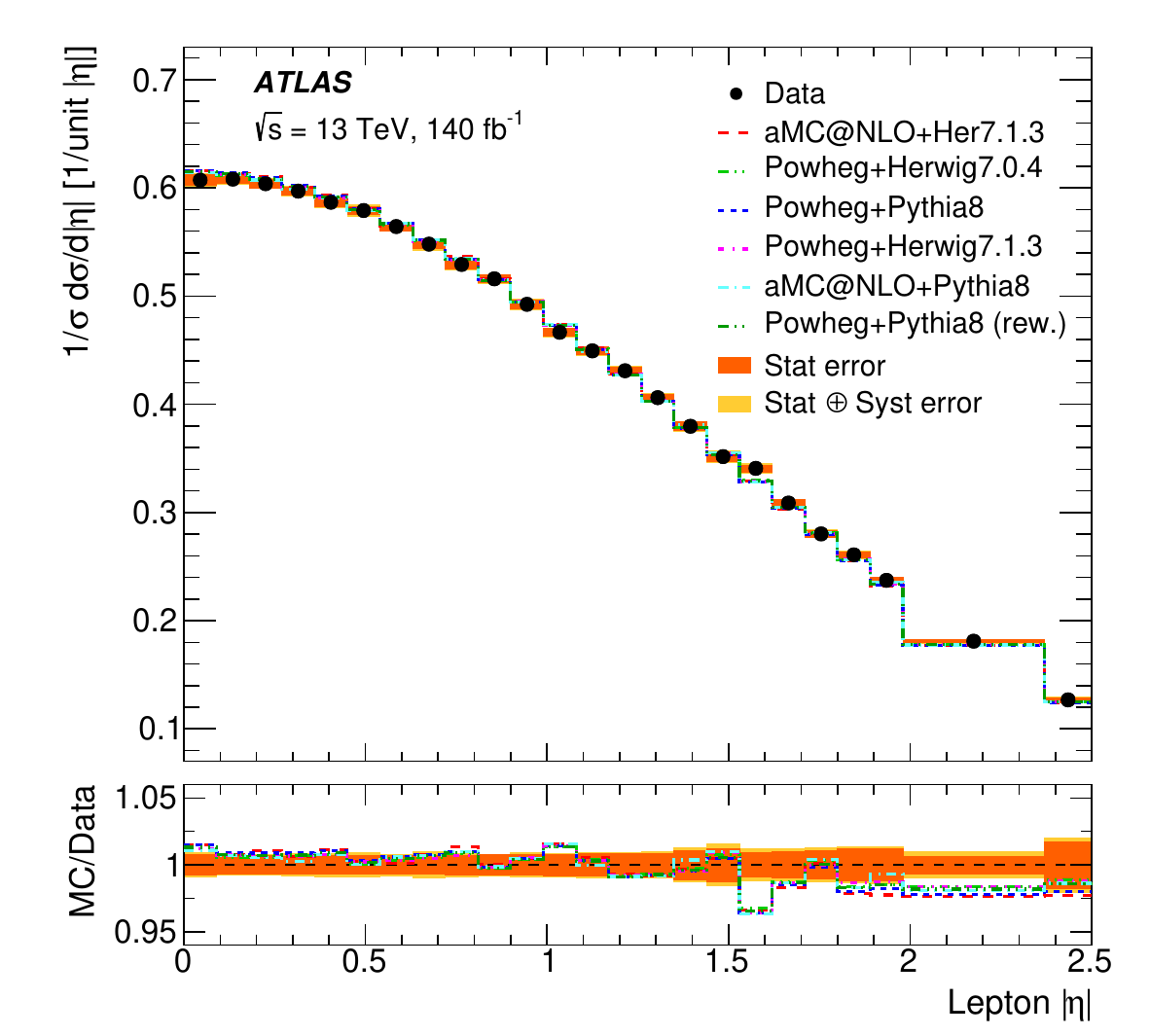} 
         \caption{}
     \end{subfigure}
     \begin{subfigure}[b]{0.32\textwidth}
         \centering
         \includegraphics[width=\textwidth]{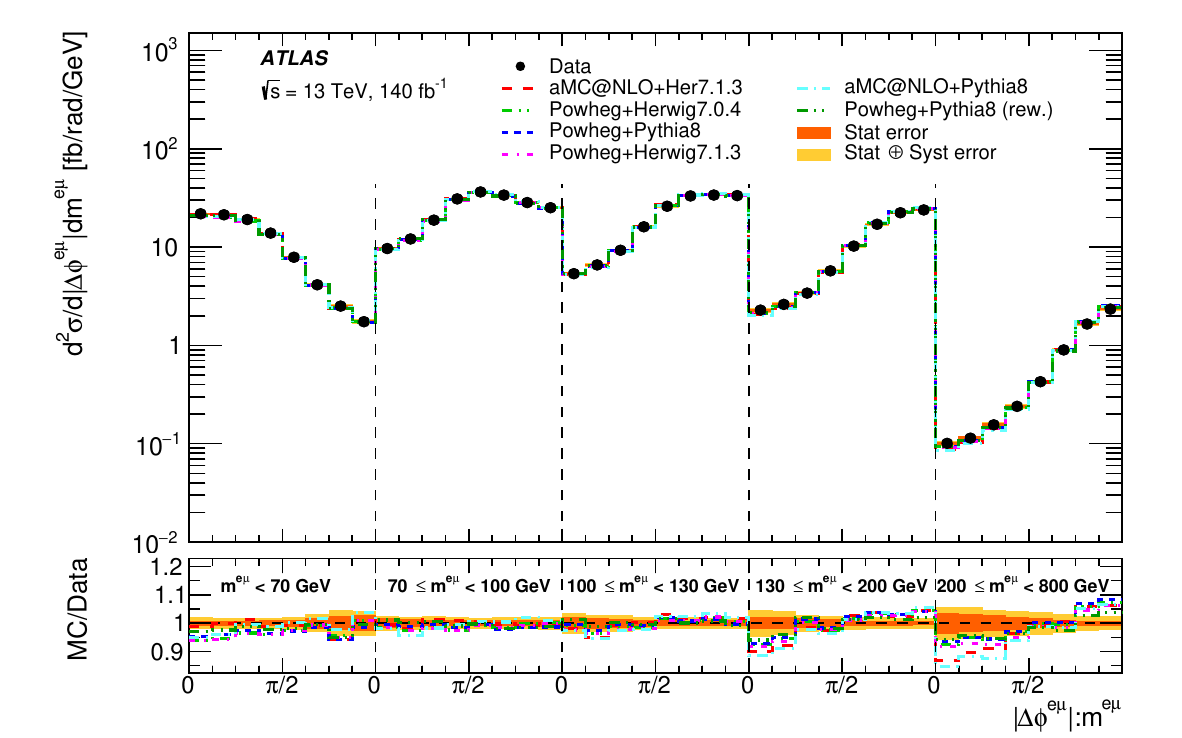}
         \caption{}
     \end{subfigure}
        \caption{\small (a) Absolute single differential cross section as a function of the transverse momentum of the lepton. (b) Normalised single differential cross section as a function of the absolute value of the pseudorapidity of the lepton. (c) Absolute double differential cross section as a function of the azimuthal difference between leptons in different regions of the invariant mass of the $e\mu$ system~[\protect\hyperlink{bib:2303.15340}{12}]. In these figures, the data are represented as dots and several NLO predictions are presented as lines in different colours.}
        \label{fig:differential leptonic}
\end{figure}

Measurements of lepton kinematic distributions were performed for $t\bar{t}$ processes selected in the $e\mu$ decay channel in $pp$ collisions at $\sqrt{s} = 13$~TeV with $\mathcal{L}=139$~fb$^{-1}$~[\hyperlink{bib:2303.15340}{12}]. Absolute and normalised differential cross sections were measured. Figure \ref{fig:differential leptonic} shows the cross sections as functions of the transverse momentum and pseudorapidity of the lepton and the double differential cross section as a function of the azimuthal difference between the two leptons in different regions of their invariant mass. The precision of the measurements is 2-3\% for the absolute cross sections and at $\approx1\%$ level for the normalised cross sections. The NLO QCD predictions do not describe all the measured observables simultaneously.\medskip

\begin{figure}[t]
     \centering
     \begin{subfigure}[b]{0.49\textwidth}
         \centering
         \includegraphics[width=\textwidth]{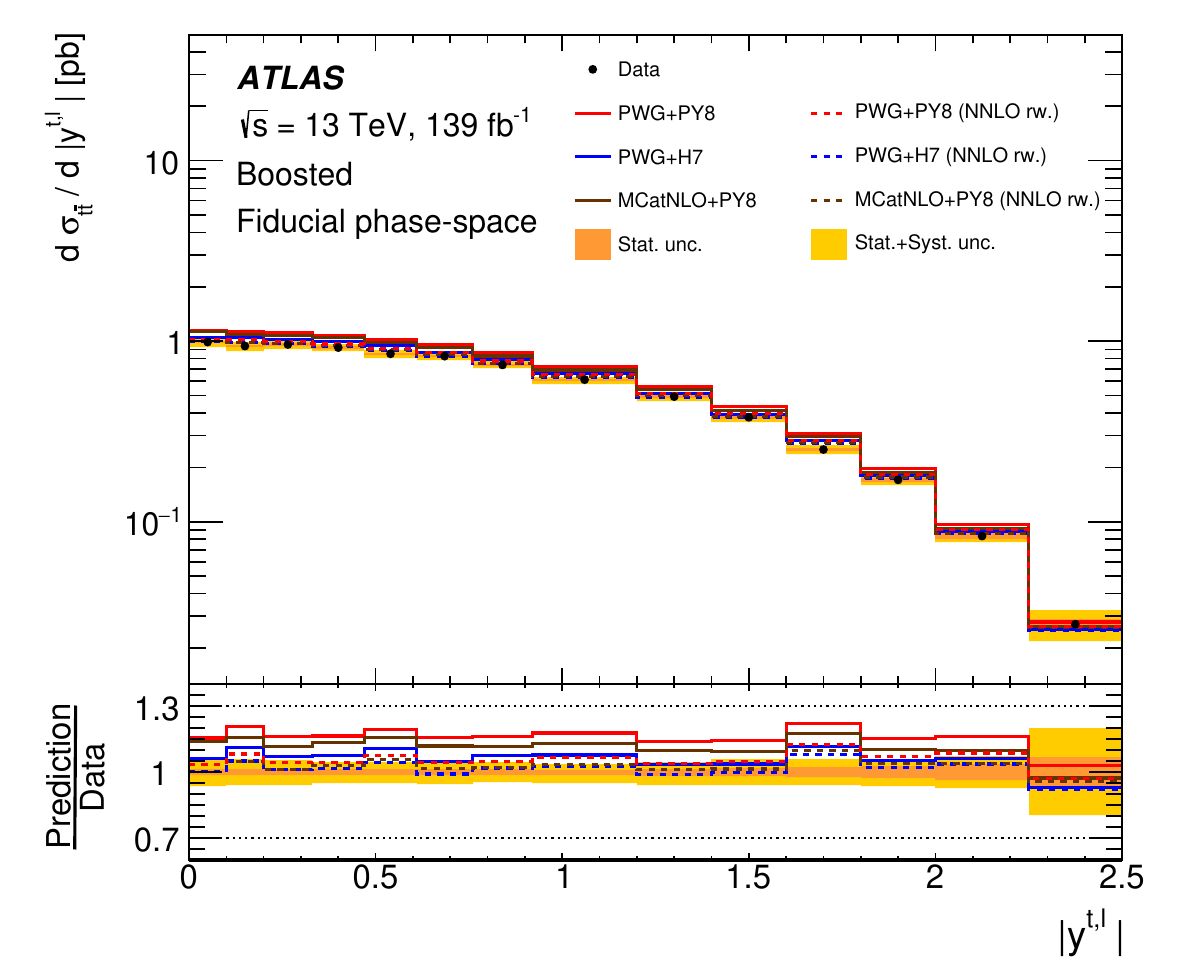} 
         \vspace{0pt}
         \caption{}
     \end{subfigure}
     \begin{subfigure}[b]{0.49\textwidth}
         \centering
         \includegraphics[width=\textwidth]{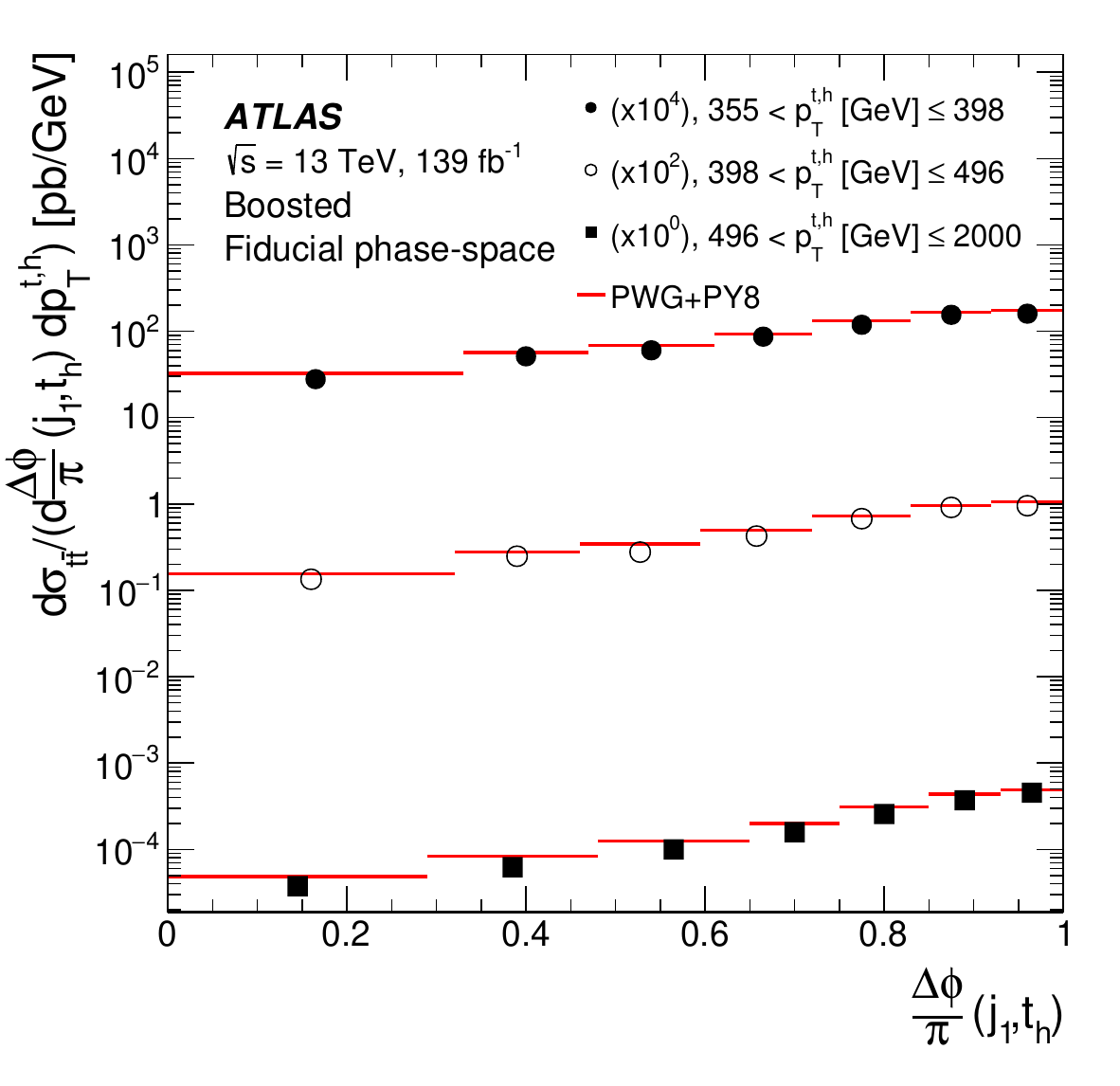} 
         \caption{}
     \end{subfigure}
        \caption{\small (a) Absolute single differential cross section as a function of the absolute value of the rapidity of the leptonic top quark. (b) Absolute double differential cross section as a function of the azimuthal difference between the hadronic top quark and the leading additional jet in different regions of the transverse momentum of the hadronically decaying top quark~[\protect\hyperlink{bib:JHEP06}{13}]. In these figures, the data are represented by markers and several NLO QCD predictions are presented as lines in different colours.}
        \label{fig:differential semileptonic}
\end{figure}

The measurement of differential $t\bar{t}$ cross sections with a high transverse momentum top quark at $\sqrt{s} = 13$~TeV with $\mathcal{L}=139$~fb$^{-1}$~[\hyperlink{bib:JHEP06}{13}] is based on the semileptonic channel; an hadronically decaying top quark was reconstructed as a $R=1$ jet with transverse momentum above $355$~GeV. This analysis introduces a novel method which uses the reconstructed top-quark mass to reduce the impact of uncertainties from the jet energy scale by introducing an scale factor. The use of this method improves significantly the precision of the measurements. The results include single and double differential cross sections; Figure \ref{fig:differential semileptonic} shows the single differential cross section as a function of the rapidity of the leptonically decaying top quark and the double differential cross section as a function of the difference in azimuthal angle between the hadronic top quark and the leading additional jet in different regions of the transverse momentum of the hadronically decaying top quark. No single prediction describes all the measured observables simultaneously. Next-to-next-to-leading-order predictions at parton level were used to reweight the NLO predictions and provide a comparison with the data; this comparison shows that the NNLO corrections are important given the precision of the measurements. An EFT interpretation of the results of this analysis was performed using \textsc{SMEFT@NLO} and EFT\textit{fitter}. Two Wilson coefficients in the Warsaw basis~[\hyperlink{bib:WARSAW}{14}] are individually and simultaneously fitted using the transverse momentum of the hadronically decaying top-quark distribution. Some of these fits provide tighter bounds than in the global fit~[\hyperlink{bib:EFT}{15}].\medskip

\textbf{5. Conclusion}\medskip

Several analyses of the ATLAS Collaboration using the datasets of the Runs 1, 2 and 3 of the LHC were presented; these include measurements of single-top and $t\bar{t}$ production at different centre-of-mass energies and in different decay channels. A combination with CMS data is also presented. From the results presented it may be concluded that leptonic channels offer a higher precision, though the novel methods developed reduce the uncertainties coming from jets. When comparing measurements with theoretical predictions it is observed that including NNLO QCD corrections improves the agreement with the data. Two BSM searches were performed using EFT interpretations with no evidence of new physics; tighter bounds for some Wilson coefficients were obtained.\medskip

\pagebreak

\textbf{6. References}\\

\hspace*{2pt} [\hypertarget{bib:ATLAS}{1}] ATLAS Collaboration, \textit{The ATLAS Experiment at the CERN Large Hadron Collider},\\
\hspace*{39pt}\href{https://iopscience.iop.org/article/10.1088/1748-0221/3/08/S08003}{JINST \textbf{3} (2008) S08003}.\medskip

\hspace*{2pt} [\hypertarget{bib:2209.08990}{2}] ATLAS Collaboration, \textit{Measurement of single top-quark production in the s-channel in\\
\hspace*{36pt} proton$-$proton collisions at $\mathrm{\sqrt{s}=13}$~TeV with the ATLAS detector}, \href{https://link.springer.com/article/10.1007/JHEP06(2023)191}{JHEP \textbf{06} (2023) 191},\\
\hspace*{37pt} arXiv: \href{https://arxiv.org/abs/2209.08990}{\texttt{2209.08990 [hep-ex]}}.\medskip

\hspace*{2pt} [\hypertarget{bib:1603.02555}{3}] CMS Collaboration, \textit{Search for $s$ channel single top quark production in $pp$ collisions at\\
\hspace*{36pt} $\sqrt{s}=7$ and $8$ TeV}, \href{https://link.springer.com/article/10.1007/JHEP09(2016)027}{JHEP \textbf{9} (2016) 027}, arXiv: \href{https://arxiv.org/abs/1603.02555}{\texttt{1603.02555 [hep-ex]}}.\medskip

\hspace*{2pt} [\hypertarget{bib:1511.05980}{4}] ATLAS Collaboration, \textit{Evidence for single top-quark production in the $s$-channel in\\
\hspace*{36pt} proton-proton collisions at $\sqrt{s}=8$ TeV with the ATLAS detector using the Matrix Element\\
\hspace*{36pt} Method}, \href{https://www.sciencedirect.com/science/article/pii/S037026931600188X?via\%3Dihub}{Phys. Lett. B \textbf{756} (2016) 228}, arXiv: \href{https://arxiv.org/abs/1511.05980}{\texttt{1511.05980 [hep-ex]}}.\medskip

\hspace*{2pt} [\hypertarget{bib:2207.01354}{5}] ATLAS Collaboration, \textit{Measurement of the $t\bar{t}$ production cross-section in $pp$ collisions at\\
\hspace*{36pt} $\sqrt{s}=5.02$ TeV with the ATLAS detector}, \href{https://link.springer.com/article/10.1007/JHEP06(2023)138}{JHEP \textbf{06} (2023) 138},\\
\hspace*{37pt} arXiv: \href{https://arxiv.org/abs/2207.01354}{\texttt{2207.01354 [hep-ex]}}.\medskip

\hspace*{2pt} [\hypertarget{bib:2212.00571}{6}] ATLAS Collaboration, \textit{Measurement of the inclusive $t\bar{t}$ production cross section in the\\
\hspace*{36pt} lepton+jets channel in $pp$ collisions at $\sqrt{s}= 7$ TeV with the ATLAS detector using support\\
\hspace*{36pt} vector machines}, 2022, arXiv: \href{https://arxiv.org/abs/2212.00571}{\texttt{2212.00571 [hep-ex]}}.\medskip

\hspace*{2pt} [\hypertarget{bib:CONF}{7}] ATLAS Collaboration, \textit{Measurement of $t\bar{t}$ and $Z$-boson cross sections and their ratio using\\
\hspace*{36pt} $pp$ collisions at $\sqrt{s} = 13.6$ TeV with the ATLAS detector}, ATLAS-CONF-2023-006, 2023,\\
\hspace*{36pt} \textsc{url}: \href{https://cds.cern.ch/record/2854834}{\texttt{https://cds.cern.ch/record/2854834}}.\medskip

\hspace*{2pt} [\hypertarget{bib:2205.13830}{8}] ATLAS and CMS Collaborations, \textit{Combination of inclusive top-quark pair production\\
\hspace*{36pt} cross-section measurements using ATLAS and CMS data at $\sqrt{s}= 7$ and $8$ TeV}, 2022,\\
\hspace*{36pt} arXiv: \href{https://arxiv.org/abs/2205.13830}{\texttt{2205.13830 [hep-ex]}}.\medskip

\hspace*{2pt} [\hypertarget{bib:JHEP04}{9}] ATLAS Collaboration, \textit{Differential $t\bar{t}$ cross-section measurements using boosted top quarks\\
\hspace*{36pt} in the all-hadronic final state with $139$ fb$^{-1}$ of ATLAS data}, \href{https://link.springer.com/article/10.1007/JHEP04(2023)080}{JHEP \textbf{04} (2023) 080},\\
\hspace*{36pt} arXiv: \href{https://arxiv.org/abs/2205.02817}{\texttt{2205.02817 [hep-ex]}}.\medskip

[\hypertarget{bib:dim6top}{10}] D. Barducci et al., \textit{Interpreting top-quark LHC measurements in the standard-model\\
\hspace*{36pt} effective field theory}, 2018, arXiv: \href{https://arxiv.org/abs/1802.07237}{\texttt{1802.07237 [hep-ph]}}.\medskip

[\hypertarget{bib:EFTfitter}{11}] N. Castro et al., \textit{EFTﬁtter --- A tool for interpreting measurements in the context of\\
\hspace*{36pt} effective field theories}, \href{https://link.springer.com/article/10.1140/epjc/s10052-016-4280-9}{JHEP \textbf{04} (2023) 080}, arXiv: \href{https://arxiv.org/abs/2205.02817}{\texttt{2205.02817 [hep-ex]}}.\medskip

[\hypertarget{bib:2303.15340}{12}] ATLAS Collaboration, \textit{Inclusive and differential cross-sections for dilepton $t\bar{t}$ production\\
\hspace*{36pt} measured in $\sqrt{s}=13$~TeV $pp$ collisions with the ATLAS detector}, 2023,\\
\hspace*{37pt} arXiv: \href{https://arxiv.org/abs/2303.15340}{\texttt{2303.15340 [hep-ex]}}.\medskip

[\hypertarget{bib:JHEP06}{13}] ATLAS Collaboration, \textit{Measurements of differential cross-sections in top-quark pair events\\
\hspace*{36pt} with a high transverse momentum top quark and limits on beyond the Standard Model\\
\hspace*{36pt} contributions to top-quark pair production with the ATLAS detector at $\sqrt{s}=13$ TeV},\\
\hspace*{38pt} \href{https://link.springer.com/article/10.1007/JHEP06(2022)063}{JHEP \textbf{06} (2022) 063}, arXiv: \href{https://arxiv.org/abs/2202.12134}{\texttt{2202.12134 [hep-ex]}}.\medskip

[\hypertarget{bib:WARSAW}{14}] B. Grzadkowski et al., \textit{Dimension-Six Terms in the Standard Model Lagrangian},\\
\hspace*{38pt} \href{https://link.springer.com/article/10.1007/JHEP10(2010)085}{JHEP \textbf{10} (2010) 085}, arXiv: \href{https://arxiv.org/abs/1008.4884}{\texttt{1008.4884 [hep-ph]}}.\medskip

[\hypertarget{bib:EFT}{15}] \textsc{SMEFiT} Collaboration, \textit{Combined SMEFT interpretation of Higgs, diboson, and top quark\\
\hspace*{36pt} data from the LHC}, \href{https://link.springer.com/article/10.1007/JHEP11(2021)089}{JHEP \textbf{11} (2021) 089}, arXiv: \href{https://arxiv.org/abs/2105.00006}{\texttt{2105.00006 [hep-ph]}}.\medskip

\end{document}